\documentclass{iaus}
\usepackage{graphicx}

\title[IAU S276.~~Searching for terrestial planets in solar analogs] 
{Volatiles and refratories in solar analogs:\\ no terrestial planet connection}

\author[Gonz\'alez Hern\'andez et al.]   
{J. I. Gonz\'alez Hern\'andez$^{1,2}$, G.~Israelian$^{1}$, N.~C.
Santos$^{3,4}$, S.~Sousa$^{3}$, E.~Delgado-Mena$^{1}$, V.~Neves$^{3}$,
\and S. Udry$^{5}$ 
}

\affiliation{$^1$Instituto de Astrof{\'\i}sica de Canarias, C/ Via
L\'actea s/n, 38200 La Laguna, Spain \\email: {\tt jonay@iac.es}\\[\affilskip]
$^2$Dpto. de Astrof\'{\i}sica y Ciencias de la
Atm\'osfera, Facultad de Ciencias F\'{\i}sicas, Universidad
Complutense de Madrid, E-28040 Madrid, Spain\\[\affilskip]
$^3$Centro de Astrof\'isica, Universidade do Porto, Rua
das Estrelas, 4150-762 Porto, Portugal\\[\affilskip]
$^4$Departamento de F\'{\i}sica e Astronomia, Faculdade 
de Ci\^encias, Universidade do Porto, Portugal\\[\affilskip]
$^5$Observatoire Astronomique de l'Universit\'e de
Gen\`eve, 51 Ch. des Maillettes, -Sauverny- Ch1290, Versoix, 
Switzerland$^1$}

\pubyear{2011}
\volume{276}  
\pagerange{xxx--xxx}
\jname{The Astrophysics of Planetary Systems: Formation, 
Structure, and Dynamical Evolution}
\editors{Alessandro Sozzetti, Mario G. Lattanzi \& 
Alan P. Boss, eds.}
\begin{document}

\maketitle

\begin{abstract}

We have analysed very high-quality HARPS and UVES spectra
of 95 solar analogs, 24 hosting planets and 71 without 
detected planets, to search for any possible signature 
of terrestial planets in the chemical abundances of volatile
and refractory elements with respect to the solar abundances.

We demonstrate that stars with and without planets in 
this sample show similar mean abundance ratios, in particular, 
a sub-sample of 14 planet-host and 14 ``single'' solar analogs 
in the metallicity range $0.14<{\rm [Fe/H]}<0.36$. 
In addition, two of the planetary systems in this sub-sample, 
containing each of them a super-Earth-like planet with masses 
in the range $\sim 7-11$ Earth masses, have 
different volatile-to-refratory abundance ratios to what 
would be expected from the presence of a terrestial planets.

Finally, we check that after removing the Galactic chemical 
evolution effects any possible difference in mean abundances,
with respect to solar values, of refratory and volatile 
elements practically dissappears.

\end{abstract}

\firstsection 

\section{Introduction \label{intro}}

The discovery of more than 400 exoplanets orbiting 
solar-type stars by the radial velocity technique have provided 
a substantial amount of high-quality spectroscopic data 
(\cite[see e.g. Neves et al. 2009]{nev09}).

Recently, \cite{mel09} have obtained a clear trend [X/Fe] versus
$T_C$ in a sample of 11 solar twins, and claimed (see also 
\cite[Ram{\'{\i}}rez et al. 2009, 2010]{ram09,ram10}) that 
the most likely explanation to this abundance pattern is related 
to the presence of terrestial planets in the solar planetary system.

Here we summarize the analysis of very high-quality HARPS and UVES 
spectroscopic data of a sample of 95 solar analogs with and 
without planets (\cite[see Gonz\'alez Hern\'andez et al.
2010]{gon10}), with a resolving power of 
$\lambda/\delta\lambda\gtrsim$85,000 and a mean $\langle$S/N$\rangle$$\sim$850.

The stellar parameters and metallicities of the whole sample 
of stars were computed using the method described in \cite{sou08}.
The chemical abundance derived for each spectral line was
computed using the LTE code MOOG (\cite[Sneden 1973]{sne73}), 
and a grid of Kurucz ATLAS9 model atmospheres 
(\cite[Kurucz 1993]{kur93}).

\section{Metal-rich solar analogs hosting super-Earth-like planets\label{secse}}

We find no substantial differences in the abundance patterns of solar
analogs with and without planets. In particular, the slopes of the
abundance ratios [X/Fe] versus $T_C$ in two metal-rich stars, HD~1461
and HD~160691, containing each of them one super-Earth-like planet, 
with 7-11 Earth masses, have the opposite sign to what one 
would expect if the amount of refractory metals in the 
atmospheres of planet hosts would depend only on the amount of 
terrestial planets.

\begin{figure}[!ht]
\centering
\includegraphics[width=4.8cm,angle=90]{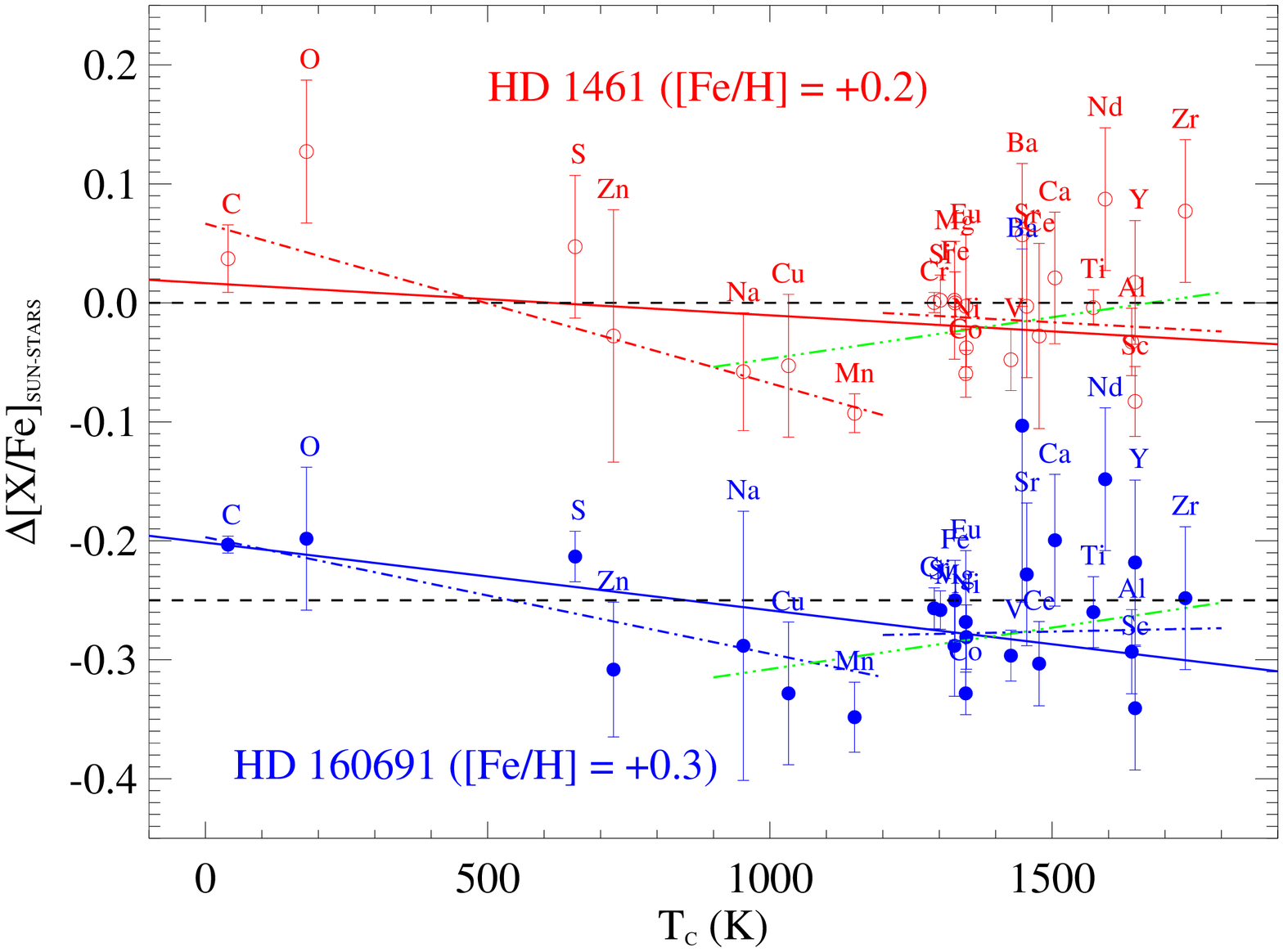}
\includegraphics[width=4.8cm,angle=90]{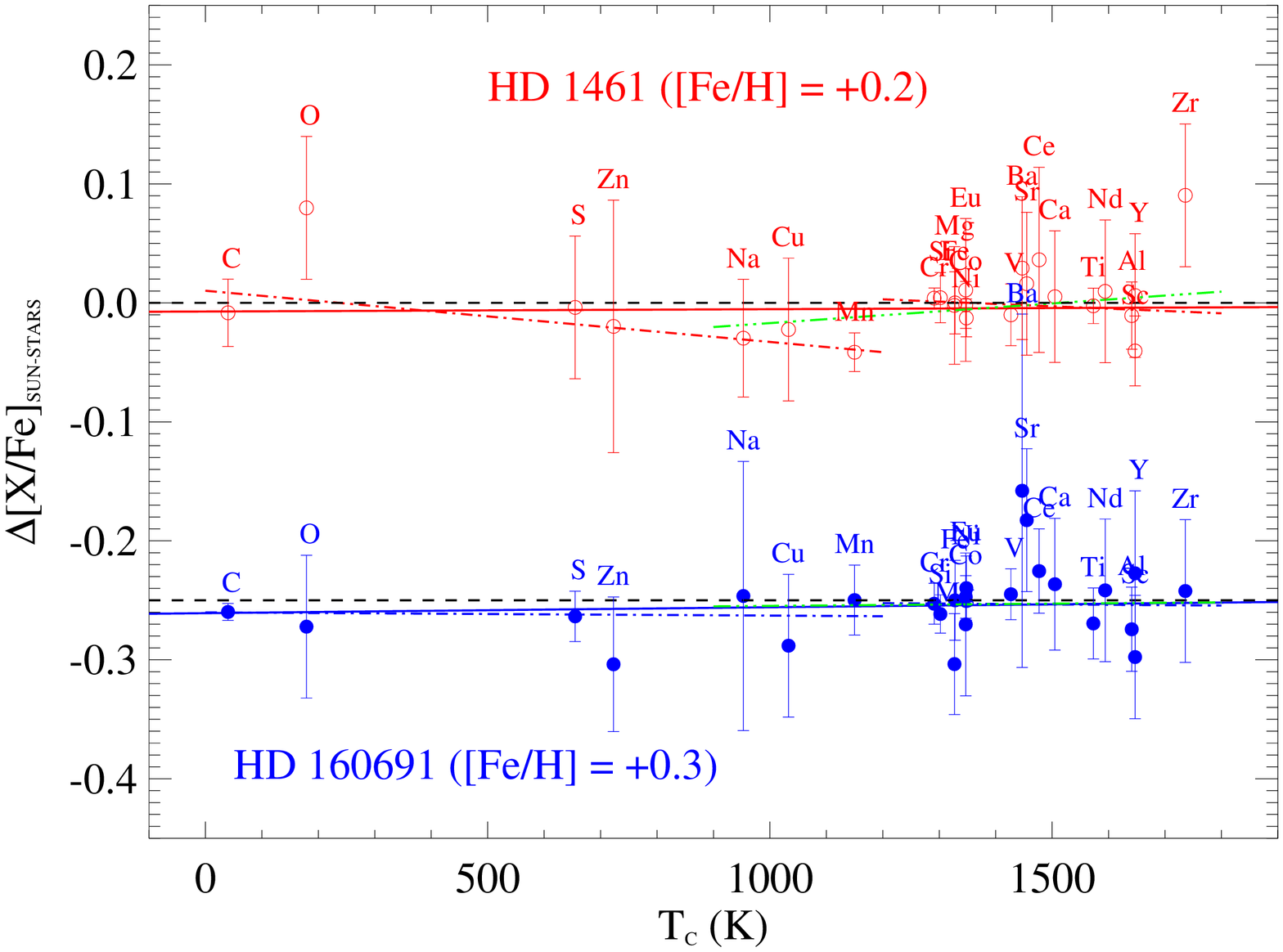}
\caption{\scriptsize{{\it Left panel}: Abundance differences, 
$\Delta {\rm [X/Fe]_{SUN-STARS}}$, between the Sun, and 
2 planet hosts with super-Earth-like planets. Linear fits for
different $T_C$ ranges to the data points weighted with the error 
bars are also displayed. We note the different slopes derived when
choosing the range $T_C > 1200$~K (dashed-dotted line) as in
\cite{mel09} and \cite{gon10}, 
and $T_C > 900$~K (dashed-three-dotted line) as in 
\cite[Ram{\'{\i}}rez et al. (2009, 2010)]{ram09,ram10}.
An arbitrary shift of -0.25~dex has been applied to the abundances 
of the planet host HD~160691.
{\it Right panel}: Same as left panel of this figure but after
correcting each element abundance ratio of each star using a
linear fit to the Galactic chemical trend of the corresponding
element at the metallicity of each star.}}   
\label{fmrse}
\end{figure}

In left panel of Fig.~\ref{fmrse} we display the abundances of 
these two stars and some linear fits for different $T_C$ ranges. 
The steep positive trend in the linear fit for $T_C > 900$~K is 
probably affected by chemical evolution effects on Mn, Na and Cu.
In right panel of Fig.~\ref{fmrse} we have 
already removed the Galactic chemical evolution effects and
both stars do not seem to show any trend. 
We may conclude that it seems plausible that many of our targets 
hosts terrestrial planets but this may not affect the
volatile-to-refratory abundance ratios in the atmospheres of these
stars (\cite[see e.g. Udry \& Santos 2007]{udr07}).

\end{document}